\documentstyle[12pt]{article}
\def\GeV{\,{\rm GeV}}
\relax

\catcode`@=11
\def\beqra{\begin{eqnarray}} \def\eeqra{\end{eqnarray}}
\def\beqast{\begin{eqnarray*}} \def\eeqast{\end{eqnarray*}}
\def\beq{\begin{equation}}      \def\eeq{\end{equation}}
\def\be{\begin{enumerate}}   \def\ee{\end{enumerate}}
\def\fo{\hbox{{1}\kern-.25em\hbox{l}}}
\def\fnote#1#2{\begingroup\def\thefootnote{#1}\footnote{#2}\addtocounter
{footnote}{-1}\endgroup}

\def\eps{\epsilon}     

\def\la{\lambda}

\def\pa{\partial}

\def\ch{\@startsection{section}{1}{\z@}{-3ex plus-1ex minus-.2ex}%
        {2ex plus.2ex}{\large\sc}}

\def\raisenot{\raise .5mm\hbox{/}}
\def\nota{\ \hbox{{$a$}\kern-.49em\hbox{/}}}
\def\notA{\hbox{{$A$}\kern-.54em\hbox{\raisenot}}}
\def\notb{\ \hbox{{$b$}\kern-.47em\hbox{/}}}
\def\notB{\ \hbox{{$B$}\kern-.60em\hbox{\raisenot}}}
\def\notc{\ \hbox{{$c$}\kern-.45em\hbox{/}}}
\def\notd{\ \hbox{{$d$}\kern-.53em\hbox{/}}}
\def\notbd{\ \hbox{{$D$}\kern-.61em\hbox{\raisenot}}} %big D
\def\note{\ \hbox{{$e$}\kern-.47em\hbox{/}}}
\def\notk{\ \hbox{{$k$}\kern-.51em\hbox{/}}}
\def\notp{\ \hbox{{$p$}\kern-.43em\hbox{/}}}
\def\notq{\ \hbox{{$q$}\kern-.47em\hbox{/}}}
\def\notW{\ \hbox{{$W$}\kern-.75em\hbox{\raisenot}}}
\def\notz{\ \hbox{{$Z$}\kern-.61em\hbox{\raisenot}}}
\def\notpa{\hbox{{$\partial$}\kern-.54em\hbox{\raisenot}}}

  \def\vk{\vec{k}}

\def\vnab{\vec{\nabla}}

\def\7#1#2{\mathop{\null#2}\limits~{#1}}        % puts #1 atop #2
\def\5#1#2{\mathop{\null#2}\limits_{#1}}        % puts #1 beneath #2

\def\lsim{\mathrel{\mathpalette\@versim<}}
\def\gsim{\mathrel{\mathpalette\@versim>}}
\def\Asl{\relax{\rm /\kern-.56em A}}
\def\Bsl{\relax{\rm /\kern-.56em B}}
\def\Dsl{\relax{\rm /\kern-.56em D}}
\def\pasl{\relax /\kern-.56em \pa}
\def\inbar{\vrule height1.5ex width.4pt depth0pt}
\def\IB{\relax{\rm I\kern-.18em B}}
\def\IC{\relax\leavevmode\hbox{\,$\inbar\kern-.3em{\rm C}$}}
\def\ID{\relax{\rm I\kern-.18em D}}
\def\IE{\relax{\rm I\kern-.18em E}}
\def\IF{\relax{\rm I\kern-.18em F}}
\def\IG{\relax\leavevmode\hbox{\,$\inbar\kern-.3em{\rm G}$}}
\def\IH{\relax{\rm I\kern-.18em H}}
\def\II{\relax{\rm I\kern-.18em I}}
\def\IK{\relax{\rm I\kern-.18em K}}
\def\IL{\relax{\rm I\kern-.18em L}}
\def\IM{\relax{\rm I\kern-.18em M}}
\def\IN{\relax{\rm I\kern-.18em N}}
\def\IO{\relax\leavevmode\hbox{\,$\inbar\kern-.3em{\rm O}$}}
\def\IP{\relax{\rm I\kern-.18em P}}
\def\IQ{\relax\leavevmode\hbox{\,$\inbar\kern-.3em{\rm Q}$}}
\def\IR{\relax{\rm I\kern-.18em R}}
\def\sed{\hbox{{\sf S}\kern-.4em\hbox{\sf S}}}
\def\ZZ{\relax{\sf Z\kern-.4em Z}}
\def\smIR{\hbox{{\footnotesize\rm I}\kern-.2em\hbox{\footnotesize\rm R}}}
\def\smIO{\ \hbox{{\footnotesize\rm I}\kern-.4em\hbox{\footnotesize\bf O}}}
\def\smIQ{\ \hbox{{\footnotesize\rm I}\kern-.5em\hbox{\footnotesize\bf Q}}}
\def\IGa{\relax{\rm I}\kern-.18em\Gamma}
\def\IPi{\relax{\rm I}\kern-.18em\Pi}
\def\IQt{\relax\leavevmode\hbox{$\kern.3em\inbar\kern-.3em\Theta$}}
\def\IOm{\relax\hbox{$\kern3.48pt\inbar\kern1.8pt\inbar\kern-5.28pt\Omega$}}
\def\ca#1{\relax\ifmmode {\cal#1} \else$\cal#1$\fi}   % Calligraphic next
\def\Sf#1{\relax\ifmmode\hbox{\sf#1}\else{\sf#1}\fi}   % San-Serif next
\def\fibby{\ifcase\@ptsize                    % redefines the Roman font
                \font\tenrm=cmfib8\or         % into the Fibonacchi font,
                \font\elvrm=cmfib8 scaled\magstephalf\or        % at the
                \font\twlrm=cmfib8 scaled\magstep1 \fi}         % @ptsize
\def\TeXey{\ifcase\@ptsize\or\or              % Instead of LaTeX's cmx12
                \font\twlrm=cmr10 scaled\magstep1     % takes 10pt fonts
                \font\twlmi=cmmi10 scaled\magstep1    % and magnifies
                \font\twlit=cmti10 scaled\magstep1    % by 120%, just
                \font\twlbf=cmbx10 scaled\magstep1\fi}  % as TeX does.

\def\ch{\@startsection{section}{1}{\z@}{-3ex plus-1ex minus-.2ex}%
        {2ex plus.2ex}{\large\sc}}
\def\sch{\@startsection{subsection}{2}{\z@}{-1.5ex plus-1ex minus-.2ex}%
        {1pt plus.2ex}{\sc}}
\def\ssch{\@startsection{subsubsection}{3}{\z@}{-1ex plus-1ex minus-.2ex}%
        {1pt plus.2ex}{\small\sc}}
\def\seceq{\@addtoreset{equation}{section}%     % Numbers Eq.s within Sect.s
        \def\theequation{\thesection.\arabic{equation}}}        % (Sect.Eq)
\def\lapp{\raisebox{-.4ex}{\rlap{$\sim$}} \raisebox{.4ex}{$<$}}
\def\gapp{\raisebox{-.4ex}{\rlap{$\sim$}} \raisebox{.4ex}{$>$}}
\def\con{\ifmmode \hbox{\bf*} \else{\bf*}\fi}   % conjugation
\def\scon{\ifmmode \hbox{\footnotesize\rm\bf*} \else{\footnotesize\rm\bf*}\fi}

\def\0#1{\relax\ifmmode\mathaccent"7017{#1}%    % puts a little circle atop,
        \else\accent23#1\relax\fi}              % as a halo of a saint

\def\place#1#2#3{\vbox to0pt{\kern-\parskip\kern-7pt
                             \kern-#2truein\hbox{\kern#1truein #3}
                             \vss}\nointerlineskip}
\def\illustration #1 by #2 (#3){\vbox to #2{\hrule width #1 height 0pt depth
0pt
                                       \vfill\special{illustration #3}}}

\def\scaledillustration #1 by #2 (#3 scaled #4){{\dimen0=#1 \dimen1=#2
           \divide\dimen0 by 1000 \multiply\dimen0 by #4
            \divide\dimen1 by 1000 \multiply\dimen1 by #4
            \illustration \dimen0 by \dimen1 (#3 scaled #4)}}

\catcode`@=12
\begin{document}
\begin{tabbing}
\hskip 11.5 cm \= {SCIPP-92-06}\\
\hskip 1 cm \> {SLAC-PUB-5740}\\
\hskip 1 cm \> {SU-ITP-92-6}\\
\hskip 1 cm \>{ }\\
\hskip 1 cm \>{February 7, 1992}\\
\end{tabbing}
\thispagestyle{empty}
\vskip 0.3cm
\begin{center}
{\Large\bf COMMENTS ON THE ELECTROWEAK}
\vskip 0.3cm
{\Large\bf PHASE TRANSITION\fnote{*}{Research supported by DOE grants
7-443256-22411-3 and DE-AC03-76SF00515, and by NSF grant PHY-8612280.
E-mail: LINDE@PHYSICS.STANFORD.EDU; DINE, HUET, LEIGH@SLACVM}}
\vskip 1.3 cm
{\bf Michael Dine and Robert G. Leigh}\\
Santa Cruz Institute for Particle Physics,\\
University of California, Santa Cruz, CA 95064\\
\vskip 0.3 cm
{\bf Patrick Huet}\\
Stanford Linear Accelerator Center\\
Stanford University, Stanford, CA 94309\\
\vskip 0.3 cm
{\bf Andrei Linde} \fnote{\dagger}{On leave from: Lebedev
Physical Institute, Moscow.}\\
Department of Physics, Stanford University, Stanford, CA 94305\\
\vskip 0.3 cm
{\bf Dmitri Linde}\\
Gunn High School, Palo Alto, CA 94305\\
\vskip 1.5 cm
{\large ABSTRACT}
\end{center}

We report on an investigation of various problems related to the
theory of the electroweak phase transition. This includes a
determination of the nature of the phase transition, a discussion
of the possible role of higher order radiative corrections and the
theory of the formation and evolution of the bubbles of the new
phase. We find in particular that no dangerous linear terms
appear in the effective potential. However, the strength of the first
order phase transition is 2/3 times less than what follows from
the one-loop approximation. This rules out baryogenesis in the
minimal version of the electroweak theory.

\vfill \newpage

1. \ \   With the recognition that baryon number violation is
unsuppressed at high temperature in the standard model has
come the realization that the electroweak phase transition
might be the origin of the observed asymmetry between matter
and antimatter \cite{Bar,mvst,ckn}. In order to have sufficient
departure from equilibrium, it is necessary that this transition
be rather strongly first order.  As a result, there has been
renewed interest in understanding under what circumstances
the transition is first order, and how the transition proceeds.
The minimal standard model almost certainly cannot produce the
observed asymmetry: it has too little CP violation, and, as we will
see, its phase transition is too weakly first order for a Higgs more
massive than the present experimentalphalimit.  Nevertheless,
for considering the features of the phase transition, it is a useful
prototype, because it is weakly coupled and comparatively simple.

Despite its apparent simplicity, understanding the phase transition
in this theory has turned out to be surprisingly complicated, and the
literature now contains contradictory claims on almost every point.
In the first papers on this subject it was assumed that the phase
transition is second order \cite{[1]}. Later it was shown
that if the Higgs mass is sufficiently small, the phase transition
becomes first order \cite{[1b]}. The problem of bubble formation
was considered in some detail; it was argued that at least initially
the bubble walls typically are rather thick \cite{[3]}.  While bubble
wall evolution was only touched upon in these early efforts
\cite{[3]}, simple arguments suggested that, e.g. for a $50$ GeV
Higgs, the motion of the wall would be non-relativistic.

Recently, quite different views have been expressed about all of
these issues.  Brahm and Hsu, in an interesting paper \cite{Hsu},
have argued that infrared effects spoil the one-loop analysis
and claim to reliably establish that the transition is
second order.  Anderson and Hall \cite{[5]} have thoroughly
considered a number of aspects of the phase transition.  Most of
their results are in qualitative agreement with the earlier
treatments \cite{[1],[1b]} and with our
previous results \cite{[4]}, while providing some quantitative
improvement.  However, they argue that the initial bubbles are thin.
Gleiser and Kolb \cite{[7]}, and Tetradis \cite{Tetr} argue that
fluctuations are so large that the transition does not proceed
through the formation of (critical) bubbles.  There has also been
controversy as to whether or not bubble wall motion is
ultrarelativistic \cite{Turok,LarryandNeil}.

In the present note, we will attempt to deal with these various
issues. We will focus on the minimal standard model.  As we have
said, this model is not realistic, but we expect
that the arguments and methods described here can
be extended to more realistic situations.  We focus on this
case only because it is the simplest -- and even here,
we will frequently have to content ourselves with rather crude
calculations. We will be able to give at least partial answers
to each of the questions raised above.  We will argue that, for
sufficiently small coupling, the phase transition can be reliably
shown to be first order. In particular, the linear term in the
potential found in Ref. \cite{Hsu} is not present.
On the other hand, we will see that the transition is more weakly
first order than suggested by the one loop analysis. We will show
that the bubbles are thick when they form, and the bubble nucleation
rate cannot be reliably computed in the thin wall approximation.
We will show that, for the range of Higgs masses considered here,
the transition does in fact proceed by nucleation of critical
bubbles.
Finally, we will make some estimates of the bubble wall thickness
and velocity.  We will see that the early estimates of Ref.
\cite{[3]} are only reliable if particle mean free paths are very long.
In practice, the relevant mean free paths are short, and the
velocity of the bubble wall somewhat larger than these early
estimates suggest.  The problem of bubble wall propagation turns out
to be surprisingly complicated, and we mention some of the issues
which must be considered.  We will see, however, that, for the
first order transitions being considered here, the
motion of the wall tends to be non-relativistic.
In this paper, we will outline our treatment of each of these
issues and describe the major results; details will be given in a
subsequent publication \cite{ourvv}.

2. \ \ We first consider the question of the order of the phase
transition. The standard approach to this problem consists of
computing the effective potential to one loop order.
At high temperature, the one-loop expression for $V_T$ is given, to a
good approximation, by
\begin{equation}\label{7}
V(\phi,T) = D (T^2 - T_o^2) \phi^2 - E T \phi^3 +
{\lambda_T\over 4} \phi^4 \ .
\end{equation}
Here $\la=m_H^2/2v_o^2$, and
\begin{equation}\label{8}
D = {1\over 8v_o^2} ( 2 m_W^2 +
m_Z^2 + 2 m_t^2) \ , \ \ E =  {1\over 4\pi v_o^3} ( 2 m_W^3 +
m_Z^3) \sim 10^{-2}\ ,
\end{equation}
\begin{equation}\label{10}
T^2_o = {1\over 4D}(m_H^2 - 8Bv_o^2) \ , \ \ B= {3\over 64\pi^2}
(2m_W^4 + m_Z^4 - 4m_t^4 ) \ ,
\end{equation}
\begin{equation}\label{11}
\lambda_T = \lambda  -  {3\over 16 \pi^2 v_o^4}
\left( 2 m_W^4 \ln{m^2_W\over a_B T^2} +
m_Z^4 \ln{m^2_Z\over a_B T^2} -
4 m_t^4 \ln{m^2_t\over a_F T^2}\right) \ .
\end{equation}
Here $v_o =246$ GeV is the value of the scalar field at the minimum
of $V(\phi,0)$ and
$\ln a_B = 2 \ln 4\pi - 2\gamma \simeq 3.51$,
$\ln  a_F = 2 \ln \pi - 2\gamma \simeq 1.14$ \cite{[1],[4],[5]}.

The behavior of $V(\phi,T)$ is reviewed in Refs. \cite{[2],sher}.
At very high temperatures the only minimum of $V(\phi,T)$ is
at $\phi = 0$. A second minimum, $\phi_1$, appears at
$T= T_1$, where
\begin{equation}\label{12}
T^2_1 = {T^2_o \over {1 - 9 E^2/8\lambda_{T_1}D}} \  , \  \
\phi_1 = {3ET\over 2\lambda_T}\ .
\end{equation}
The values of  $V(\phi,T)$ in the two minima become equal
to each other at the temperature $T_c$, where
\begin{equation}\label{14}
T^2_c = {T^2_o \over {1 -  E^2/\lambda_{T_c}D}} \  , \  \
\phi_c =  {2ET\over \lambda_T} \ .
\end{equation}
The minimum of $V(\phi,T)$ at $\phi = 0$ disappears at the
temperature $T_o$, when the field $\phi$ in the second
minimum is $\phi_o = {3ET/\lambda_T}$.

3. \ \ The first order character of the transition is due to the
$\phi^3$ term in the potential.
The appearance of such a term non-analytic in $\vert \phi \vert^2$
is the signal of an infrared problem, and raises concerns about
the validity of the perturbation expansion.  Indeed, these issues
were raised in the early work on the subject \cite{[1b],[11],LGPY}.
The problem of infrared divergences at high temperatures in
theories with light or massless particles has received extensive
attention in the literature and is discussed in many textbooks
\cite{[2],Kapusta}. It is well known that
these problems arise from the zero-frequency terms in the
discrete frequency sums which appear in the computation of
equilibrium quantities at finite temperature.
The corresponding Feynman diagrams
are those appropriate to a three dimensional field theory.
For simplicity, we will consider here the contribution of W bosons.
In the present case, the problem is most easily analyzed in Coulomb
gauge, $\vnab\cdot \vec W =0$.  The relevant propagators
are those for the Coulomb lines, $D_{00}$, and
transverse lines, $D_{ij}$.  For zero frequency,
\begin{equation}\label{i1}
D_{00} = {1 \over \vk^2} \ , \ \
D_{ij}(\vk) = {1 \over { \vk^2 + m^2_W(\phi)}}P_{ij}(\vk) \ ,
\end{equation}
where $m_W = g \phi/2$,
and $P_{ij}=\delta_{ij} -{k_i k_j \over \vk^2}$.
The 'Goldstone' field and the physical Higgs scalar
have standard scalar propagators with mass terms which are
independent of the gauge coupling.  The cubic term is readily
extracted from the zero-frequency piece of the determinant.
$2/3$ of it arises from the transverse gauge bosons; the other
$1/3$ is obtained from the Coulomb line.

In this gauge it is rather easy to see how higher orders of
perturbation theory behave.  At one loop, it is well-known that
the Coulomb line, even for $\phi=0$, acquires a mass
$m_{D}^2 = (N_g + N_c) g^2 T^2/3$ \cite{Gluon}.
In order to obtain a sensible perturbation theory for small $\phi$,
it is necessary to partially resum the perturbation expansion,
i.e. to use in each order of the loop expansion
a Coulomb propagator with this effective
mass. In the infra-red, this resummation corresponds to integrating
out all heavy modes (to one loop order),
leaving only the $\omega=0$ modes, with a Debye mass. In practice,
the tree level gauge boson mass at the minimum of the potential is
small compared to $m_{D}$.  As a result, repeating the one loop
calculation with this effective mass, the contribution to the
cubic term from the Coulomb line disappears.  The same is not true,
however, for the transverse bosons.  For zero $\phi$, gauge
invariance forbids a one-loop mass for these fields; the transverse
polarization tensor is in fact given by
$\Pi_{ij}={15 g^2T \sqrt{\vk^2} \over 32}P_{ij}$.
As a result, the contribution to the cubic term from these fields
survives, and the net effect is to reduce the coefficient of the
cubic term in eq. (\ref{7}) by $2/3$:
\footnote{After we obtained this result, we
received a very interesting paper by Carrington \cite{Carr} where the
modification of the cubic term by high order corrections was also
considered. Even though the author did not claim that these
corrections reduce the cubic
term by a factor of $2/3$, after some algebra one can
check that his result is equivalent to ours.}
\begin{equation}
E =  {1\over 6\pi v_o^3} ( 2 m_W^3 + m_Z^3) \ .
\end{equation}

This reduction means that, for a given Higgs mass, the phase transition
is significantly less first order than one expects from the one loop
analysis.  It has been pointed out that a minimal requirement of the
phase transition is that the sphaleron rate after the transition
be sufficiently small that the baryon number not be washed out.
Using the (unimproved) one loop result, this gives a limit of about
$42-55$ GeV \cite{[8],[4]}, at best just barely consistent with the
present experimental constraints \cite{LEP}. Allowing for the
correction obtained here, the limit on $m_H$ is reduced by about
$25\%$, clearly ruling out baryogenesis in the model.

More generally, however, we can ask about the behavior of the
perturbation
expansion, particularly at small $\phi$.  Before making general
remarks, it is helpful to consider the two loop diagrams involving
transverse gauge bosons and scalars (Fig. 1).  The zero frequency
pieces of these diagrams, separately,
 give contributions $\sim g^3 \vert \phi \vert T^3$.  If
one combines these diagrams, however, being careful
about combinatorics, they have the structure of an
insertion of a polarization tensor on the transverse gauge boson
line.  Because, as mentioned above, this tensor vanishes
at zero momentum, the sum of these diagrams is less singular at
small $\phi$, and simply gives a
correction to the quadratic term.  We have checked all other
diagrams at two loop order and shown that there are no linear
terms in the potential.

The authors of Ref. \cite{Hsu} found a linear
contribution to the potential by simply substituting the
scalar mass found at one loop back into the one loop
calculation.  Such a procedure is generally reliable when
calculating Green's functions or tadpoles.
Indeed, it is well known that the sum of the geometric progression,
which appears after the insertion of an arbitrary number of
polarization operators $\Pi(T)$ into the propagator
$ (k^2 + m^2)^{-1}$, simply
gives  $(k^2 + m^2 + \Pi(T))^{-1}$. However, this trick does
not work for the closed loop diagram for the effective potential,
which contains $\ln (k^2 + m^2)$. A naive substitution of the
effective mass squared $m^2 + \Pi(T)$ instead of $m^2$ into
$\ln (k^2 + m^2)$ corresponds to a wrong counting of
higher order corrections.
The simplest way to avoid the ambiguity is to calculate tadpole
diagrams for ${dV\over d\phi}$ instead of the vacuum loops,
and then integrate the result with respect to $\phi$. One can easily
check by this method as well,  that  no linear terms appear
in the expression for  $V(\phi,T)$.

The absence of linear terms does not automatically mean that higher
order corrections are completely under control.
A general investigation of the infrared problem in the non-Abelian
gauge theories at finite temperature suggests  that the results
which we obtained are reliable for $\phi \, \gapp \, {g\over 2}\, T
\sim T/3 $ \cite{[11],LGPY}.
Thus, a more detailed investigation is needed to study behavior
of the theories  with $m_H \,\gapp\, 10^2$ GeV near the critical
temperature,  since the scalar field, which appears at the moment
of the phase transition in these theories,
is very small (see Fig. 2). However,  we expect that our results are
reliable for strongly first order phase transitions with
$\phi \,\gapp\, T$, which is quite sufficient to study (or to rule
out) baryogenesis in the electroweak theory.

4. \ \ We turn now to the problem of bubble formation.  At high
temperatures,
this becomes a problem in classical thermodynamics.  One looks for
a stationary point of the free energy, with the property that
$\phi \rightarrow 0$ as $r \rightarrow \infty$, i.e. a solution
of the classical field equations with potential $V_0 + V_T$.
Some care is required in solving this equation, however, since
necessarily one is constructing a saddle point of the action
(unstable modes corresponding to bubble growth or collapse).
As a result, if one makes a poor approximation, one overestimates
the probability of bubble formation.  We believe this is the case
of the analysis of Ref. \cite{[5]}, where formation of bubbles was
studied in the thin wall approximation.  This approximation works
well if the height of the barrier between the two minima is much
larger than the difference between
the values of the effective potential in each of them. This is not
the case for the phase transition with $m_H \,\lapp\, 60$ GeV.
Numerical solution of the equations yields an action typically a
few times larger than that obtained from the thin wall approximation.

Even though the thin wall approximation fails, one can still
study bubble formation analytically in a wide class of theories.
Indeed, we have found that in the vicinity of the phase transition
one can write, to a very good approximation,
\begin{equation}\label{30}
{S_3\over T} =  {38.8\,D^{3/2}\over {{{ E}^2}}}\cdot \left({\Delta
T\over T}\right)^{3/2} \times f\Bigl({2\,\lambda_o\,D\,\Delta T\over
E^2\,T}\Bigr).
\end{equation}
where
\begin{equation}\label{29}
f(\alpha) = 1 + {\alpha\over 4} \Bigl[ 1
+{{2.4}\over {1 - \alpha}} +
 {{0.26}\over{{{( 1 - \alpha ) }^2}}}\Bigr] .
 \end{equation}
Parameter $\alpha$ changes from $0$ to $1$ in the range of
temperatures for which the phase transition is possible. We have
found that eq. (\ref{29}) is correct with an accuracy about $1\%$
in the most interesting range $0 \leq \alpha \leq 0.95$.

5. \ \ In the case that the transition is weakly first order, it is
natural to ask whether the transition actually proceeds through
formation of bubbles, or if other sorts of fluctuations might be more
important. In most treatments, it is assumed that the transition
occurs once the bubble
nucleation rate is large enough that the universe can fill with
bubbles.  In practice, because of the extremely slow expansion
rate at the time of the transition, this means that the barrier is
still high
enough that the naive calculation of the nucleation rate gives
an extremely small result; the three dimensional action is of order
$130 - 140$. Given that the rate of formation of
critical bubbles is so small, one might expect that other types
of fluctuations which might equilibrate the two phases would be
extremely rare. In Refs. \cite{[7],[12],Tetr},
however, it has been argued that this is not the case.
Roughly speaking, these authors arrive at this conclusion by estimating
the mean square fluctuation of the scalar field about the symmetric
minimum, $\phi^2_{rms} = <\phi^2>$, and comparing
this with the value of the field at the other
minimum.  A rough estimate leads them to the conclusion that the
$<\phi^2> \sim mT \sim \phi^2_c$, so that it is not meaningful to
consider the system as sitting in one vacuum or the other.
Here $m$ is the Higgs field mass near $\phi = 0$. Subcritical
bubbles, they argue, equilibrate the two phases even before one
reaches the temperature $T_c$.

While we believe that for some range of parameters subcritical
bubbles may be important, we do not believe that this is the case
for the Higgs masses under consideration here.
In estimating $\phi_{rms}$, one should be careful to
consider only long wavelength modes.  Short wavelength
modes will be associated with configurations with large gradient
terms, which will collapse in a microscopic time.  Also, one must
be very careful with factors of $\pi$ and $2$.  A more detailed
investigation based on the stochastic approach to tunneling
gives the estimate for the amplitude of relevant fluctuations
$<\phi^2> \sim {mT\over \pi^2}$ \cite{[10]}.
Combining this estimate  with  our results for the mass $m$,
i.e. for the curvature of the effective potential near
$\phi = 0$ in the relevant temperature interval,
$T_o < T < T_1$, one obtains an estimate $\phi_{rms} \sim .1 \;T$.

{}From this analysis, we see that the typical amplitude
of the relevant scalar field fluctuations is substantially less
than the separation of the two minima, unless the phase transition
is very  weakly first order. Even for $m_H \sim 60$ GeV,
 the distance between the two minima remains five times greater
than $\phi_{rms}$.
Including fluctuations with $k \gg k_{max}$ will give larger
$\phi_{rms}$, but these will collapse in a microscopic time, and
will not serve to equilibrate the two phases.

6. \ \
Understanding the motion of the wall is important mainly in the
context of baryogenesis.  We have already seen that no baryons
will be generated in the single Higgs theory, unless the Higgs mass
is well below the present experimental limit.  We expect, however,
that, for $m_H \sim 35$ GeV (for which $\phi /T \, > \, 1$),
the bubble wall motion in this model will have many features in
common with more realistic theories of baryogenesis, which require
the phase transition to be strongly first order.
Even in this simple model, determining the wall velocity and shape
is a difficult problem, and we will content ourselves with rather
crude estimates. In our analysis, we will assume
that the wall achieves a steady state after
some time. In particular, there is a frame, which we refer to as the
`wall frame',  in which the scalar field and the particle
distributions are independent of time.
We will assume that the principle source of damping of the walls
motion is elastic scattering of particles
from the wall; if there are additional sources of damping, they
can only slow the wall even further.  We will treat the
velocity of the wall as a small parameter.
We will have to check its validity {\em a posteriori}.
With these assumptions, it is helpful to
consider two limiting cases, depending on whether the size of
the wall, $\delta$, is large or small compared to the relevant mean
free paths for elastic scattering, $\zeta$.  As an estimate of these
mean free paths, we follow Ref. \cite{ckn} and take the longitudinal
and transverse gluon propagators to include a mass proportional to
$m_{D}$. For top quarks, this yields
$\sigma_t = {16 \, \pi \alpha_s^2 / 3 m_{el}^2}$.
Multiplying by the flux one obtains $\zeta \sim 4\, T^{-1}$
for quarks.  For $W$'s and $Z$'s the result is about three times
larger. These numbers are consistent with results which have been
obtained for the stopping power \cite{BraThom}.
To get some feeling for the wall size, consider the system at
temperature $T_c$.  At this temperature, the two phases can
coexist, separated by a static domain wall.  The $\phi$ field
in this domain wall is readily obtained by quadrature; for a
$35 \GeV$ Higgs, one finds $\delta\sim 40\, T^{-1}$.

This suggests that the thin wall limit may not be a good one, but it
is instructive to consider it in any case. In this limit, a typical
particle passes through the
wall, or is reflected from it, without scattering.  An estimate
in this limit was given in Ref. \cite{[3]}.  If the problem
is treated semiclassically, it is straightforward
to calculate the extra, velocity-dependent force
on the wall, assuming that all particles approaching the wall from
either side
are described by an equilibrium distribution at some temperature.
(Note that particles moving away from the wall are not described
by an equilibrium distribution in this case; these particles are
assumed to be equilibrated far from the wall, at some possibly
different temperature and velocity).  In this model it is
straightforward to calculate the force on the wall to linear
order in $v$. For bosons, the leading term is of order $m^3T$, while
for fermions, it is of order $m^4$. Equating this force to the
pressure difference on the two sides of the wall gives the velocity.
Defining
\beq\label{paraexp}
\eps = {T_c- T \over T_c - T_o} \ ,
\eeq
we can obtain an approximate equation for the velocity, valid for
small $\epsilon$:
\beq
v\sim\frac{\pi}{6}\frac{\eps}{(1+\eps)}\; {\cal C}(m_t,\la) \ ,
\label{eq:appr}
\eeq
where the correction ${\cal C}$ comes predominantly from top quarks and
is roughly of order one. For $m_t=120$ GeV and $m_H = 35$ GeV (for
which $\eps\sim 1/4$), we find $v\sim0.05$.

In this discussion we have made a variety of oversimplifications.
One which is potentially important is our assumption of equilibrium
densities on both sides of the wall.  Some fraction of
incoming particles, however, are reflected from the wall, and this
will tend to lead to an enhancement of the particle density in front
of the wall.  For definiteness, consider top quarks.
As a crude estimate, suppose the fraction of reflected particles is
$f$ (of order $m/T$), and
suppose that the mean free path for processes which can change
top quark number is $\tau$, with $\zeta \ll \tau$.
Then the density in front of the wall is enhanced by an amount of
order $fvn\sqrt{{\tau\over\zeta}}$. For the thin wall case,
this is likely to reduce the velocity of the wall somewhat.
In the thicker wall cases considered below, however, this is likely
to be more important.

Consider now the case $\delta\gg\zeta$.
In this case, to model the deviations from equilibrium due to the
finite velocity of the wall, we can break the wall into segments
of length $\zeta$, and repeat the thin wall analysis for each
of these segments.  In particular, we assume that the distributions
on either side of each segment are at equilibrium.
We also assume that the temperature and velocity of the
particle distributions are constant across the wall.  This should
not be a bad approximation when there are many light species of
particles
in the system.  One obtains for the $v$-dependent force on the
wall, a result roughly suppressed by a factor of order $\sqrt{\zeta
/\delta}$. For a Higgs mass of $35$ GeV we obtain a velocity $\sim
0.2$.

In this thick wall case, the density enhancement described
earlier could be extremely important.  Here one expects the
extra particles to be distributed more or less uniformly
over the wall.  This leads to an extra force
$\Delta F \sim fnv\tau\; \Delta \rho\; /\delta$
where $\Delta \rho$ denotes the internal energy difference on the two
sides of the wall.  As an estimate of $f$, we can take the ratio
of the equilibrium densities on the two sides of the wall; for top
quarks, this gives a number
of order $5\%$.  This effect seems to be comparable to that of the
preceding paragraph.

7. \ \ In this paper we have outlined a program for treating the
phase transition in weakly coupled theories with scalar fields.
 From the standpoint of electroweak baryogenesis,
it is important to find
models consistent with current experimental bounds in which
the phase transition is strongly first order.
It would be interesting to find models where the bubble wall
is thin; in such theories \cite{ckn} electroweak baryogenesis
can be very efficient.  However, our work suggests that
in many models, the wall will be slow and thick, and adiabatic
analyses of the type of Ref. \cite{mvst} will be relevant.  In these
cases, one may just barely be able to produce the observed asymmetry.
Further improvements in both the theory of the phase transition
and that of electroweak baryogenesis will be necessary to completely
settle these questions.

We are grateful to L. Susskind, L. McLerran, R. Kallosh,
N. Turok and M. Gleiser for discussions.

\pagebreak


\begin{thebibliography}{999}
\bibitem{Bar} V.A. Kuzmin, V.A. Rubakov and M.E.Shaposhnikov, Phys.
Lett.
{155B} (1985) 36; M.E. Shaposhnikov, JETP Lett. {\bf 44} (1986) 465;
Nucl. Phys. {\bf B287} (1987) 757;  Nucl. Phys. {\bf B299} (1988)
797; A.I. Bochkarev, S.Yu. Khlebnikov and M.E. Shaposhnikov,
Nucl. Phys. {\bf B329} (1990) 490;
 L. McLerran, Phys. Rev. Lett. {\bf 62} (1989) 1075;
 N. Turok and P. Zadrozny, Phys. Rev. Lett.
{\bf 65} (1990) 2331; Nucl. Phys. {\bf B358} (1991) 471.
\bibitem{mvst}  L. McLerran, M. Shaposhnikov, N. Turok and
M. Voloshin, Phys. Lett. {\bf 256B} (1991) 451;
 M. Dine, P. Huet, R. Singleton and L. Susskind,
Phys.Lett. {\bf 257B} (1991) 351.
\bibitem{ckn}
 A. Cohen, D.B. Kaplan and A.E. Nelson, Nucl. Phys.
 {\bf B349} (1991) 727; Phys.Lett.
{\bf 263B} (1991) 86;  University of
California, San Diego, preprint  UCSD-PTH-91-20 (1991).
\bibitem {[1]} D.A. Kirzhnits, JETP Lett. {\bf 15} (1972) 529;
 D.A. Kirzhnits and  A.D. Linde, Phys. Lett. {\bf 72B} (1972) 471;
  S. Weinberg, Phys. Rev. {\bf D9} (1974) 3357;
  L. Dolan and R. Jackiw, Phys. Rev. {\bf D9} (1974) 3320;
 D.A. Kirzhnits and  A.D. Linde,
JETP {\bf 40} (1974) 628.
\bibitem {[1b]} D.A. Kirzhnits and  A.D. Linde, Ann. Phys. {\bf 101}
(1976) 195.
\bibitem {[3]} A.D. Linde, Phys.Lett. {\bf 70B} (1977) 306; {\bf 100B}
(1981) 37; Nucl. Phys. {\bf B216} (1983) 421.
\bibitem{Hsu} D. Brahm and S. Hsu, Caltech
preprints CALT-68-1705 and CALT-68-1762 (1991).
\bibitem{[5]} G. Anderson and L. Hall, LBL preprint
LBL-31169 (1991).
\bibitem {[4]} M. Dine, P. Huet and R. Singleton,   University of
California, Santa Cruz, preprint SCIPP 91/08 (1991);
A.D.  Linde and D.A. Linde, unpublished.
\bibitem {[7]} M. Gleiser and E. Kolb, FNAL preprint
FERMILAB-Pub-91/305-A (1991).
\bibitem{[12]}M. Gleiser, E. Kolb and R. Watkins, Nucl. Phys.
{\bf B364} (1991) 411.
\bibitem{Tetr} N. Tetradis, preprint DESY 91-151.
\bibitem {Turok} N. Turok, Princeton preprint PUPT-91-1273.
\bibitem{LarryandNeil}L. McLerran and N. Turok, private
communication.
\bibitem{ourvv} M.Dine, R.G. Leigh, P. Huet,
A.D. Linde and D.A. Linde,
 preprint SCIPP-92-07, SLAC-PUB-5741, SU-ITP-92-7 (1992).
\bibitem {[2]} A.D. Linde, {\em Particle Physics and Inflationary
Cosmology} (Harwood, New York, 1990).
\bibitem {sher} M. Sher, Phys. Rep. {\bf 179} (1989) 273.
\bibitem {[11]} A.D. Linde, Rep. Prog. Phys. {\bf 42} (1979) 389.
\bibitem{LGPY} A.D. Linde, Phys. Lett. {\bf 93B} (1980) 327; \\
D.J. Gross, R.D. Pisarski and L.G. Yaffe, Rev. Mod. Phys. {\bf 53}
(1981) 1.
\bibitem{Kapusta} J.I. Kapusta, {\em Finite Temperature Field
Theory}, Cambridge University Press, 1989.
\bibitem {Gluon} O.K. Kalashnikov and V.V. Klimov,
Sov. J. Nucl. Phys. {\bf 31} (1980) 699;
H.A. Weldon, Phys. Rev. {\bf D26} (1982) 1384, 2789;
E. Braaten and R. Pisarski, Phys. Rev. {\bf D42} (1990) 2156.
\bibitem{Carr} M.E. Carrington, University of Minnesota preprint
TPI-MINN-91/48-T (1991).
 \bibitem{[8]} A. Bochkarev, S. Kuzmin and M. Shaposhnikov, Phys.
Lett. {\bf 244B} (1990) 27.
\bibitem{LEP} ALEPH, DELPHI, L3 and OPAL Collaborations,
as presented by M. Davier, Proceedings of the International
Lepton-Photon Symposium and Europhysics Conference on
High Energy Physics, eds. S. Hegerty, K. Potter and E. Quercigh
(Geneva, 1991), to appear.
\bibitem {[10]} A.D. Linde, Stanford University preprint SU-ITP-900
(1991), to be published in Nucl. Phys.
\bibitem {BraThom} E. Braaten and M.H. Thoma, LBL
preprint LBL-30998 (1991).

\end{thebibliography}
\end{document}